\documentclass[fleqn,usenatbib,useAMS]{mnras}

\usepackage{graphicx}	
\usepackage{amsmath}	
\usepackage{multicol}        
\usepackage{bm}		
\usepackage{pdflscape}	
\usepackage{mhchem} 

\usepackage[T1]{fontenc}
\usepackage{ae,aecompl}
\usepackage[normalem]{ulem}
\usepackage{newtxtext,newtxmath}

\usepackage{natbib}
\defcitealias{Molero2025}{M25}

\title[\texorpdfstring{Modelling chemical clocks with Kepler}{Modelling [s/alpha]}]{Modelling s-process chemical clocks: insights from high-precision \emph{Kepler} data}

\author[G. Casali \& M. Molero]{
G. Casali$^{1,2}$\thanks{These authors contributed equally to this work.}
\thanks{E-mail: giada.casali@anu.edu.au} and
M. Molero$^{3,4}$\footnotemark[1]
\thanks{E-mail: marta.molero@tu-darmstadt.de} 
\\
$^{1}$Research School of Astronomy and Astrophysics, The Australian National University, Canberra, ACT 2611, Australia\\
$^{2}$INAF -- Osservatorio di Astrofisica e Sceinza dello Spazio, Via P. Gobetti, 93/3, 40129, Bologna, Italy\\
$^{3}$Institut für Kernphysik, Technische Universität Darmstadt, Schlossgartenstr. 2, Darmstadt 64289, Germany\\
$^{4}$INAF -- Osservatorio Astronomico di Trieste, Via Tiepolo 11, I-34131 Trieste, Italy\\
}
\date{Accepted 2026 August 11. Received 2026 July 21; in original form 2026 June 16}

\pubyear{{\the\year{}}}

\begin{document}
\label{firstpage}
\pagerange{\pageref{firstpage}--\pageref{lastpage}}
\maketitle

\begin{abstract}
We present Galactic chemical evolution (GCE) models for the chemical clocks [Zr/Ti] and [Ce/Ti], tracing first- and second-peak s-process nucleosynthesis, and compare them with a high-precision sample of 68 \emph{Kepler} red giant stars with asteroseismic ages from individual-mode frequencies and high-resolution spectroscopy. Using a multi-zone GCE framework, we explore variations in metallicity-dependent asymptotic giant branch (AGB) nucleosynthetic yields, including proposed enhancements to high-metallicity Ce production.
Our baseline model reproduces [Zr/Ti] and the high-$\alpha$ sequence in both age and metallicity space, but systematically underestimates [Ce/Ti] at young ages and intermediate metallicities, indicating a persistent deficit in second-peak s-process enrichment over the last $\sim6$ Gyr of Galactic disc evolution.
Increasing second-peak yields from high-metallicity AGB stars only partially reduces this discrepancy, suggesting that simple yield rescaling is insufficient and more fundamental revisions to s-process nucleosynthesis at high metallicity, alongside a self-consistent treatment of stellar dynamics, may be required. In fact, models reproduce abundance trends more tightly in metallicity than in age space, with additional age scatter partly attributed to radial migration.
This Letter highlights the diagnostic power of precise asteroseismic ages for GCE studies and the limitations of current models in capturing the complex interplay between s-process nucleosynthesis and stellar dynamics.  


\end{abstract}





\section{Introduction}

The ratio of slow neutron-capture (s-process) elements to $\alpha$-elements, [s/$\alpha$], is a promising potential stellar age indicator, a so-called \textit{chemical clock}. The physical basis of this approach lies in different nucleosynthetic timescales of these two classes of elements: $\alpha$-elements are produced primarily in core-collapse supernovae (CC-SNe) on short timescales, whereas s-process elements are synthesised in low-intermediate mass stars (LIMS) during their asymptotic giant branch (AGB) phase, over several Gyr, with a further contribution to the weak s-process from massive stars. Their ratio, therefore, evolves with time, generally increasing towards younger stellar ages, and has been successfully used to assign statistical ages to large populations of disc stars (e.g., \citealp{spina18,delgado19,casali25} and references therein).

Nevertheless, a growing body of evidence demonstrates that the age-[s/$\alpha$] relation is not universal. Its slope depends on the specific combination of s-process and $\alpha$-elements, on the stellar metallicity, and on the location within the Galactic disc (e.g., \citealp{feltzing17,casali20,viscasillas22,salessilva22}). Open clusters from the \emph{Gaia}-ESO survey \citep{randich22}, spanning a wide range of ages and Galactocentric distances, have shown that chemical clocks are steeper in the inner disc and for the second s-process peak elements (Ba, La, Ce), questioning whether these spatial and elemental variations are at least partly driven by the metallicity dependence of AGB nucleosynthetic yields (\citealp{magrini21,casali23,Molero2025}, hereafter \citetalias{Molero2025}). Furthermore, \citet{Ratcliffe2024} demonstrated, using estimated stellar birth radii for a large sample of disc stars observed by APOGEE \citep{apogeedr17} and GALAH \citep{galah21} surveys, that the non-universality of chemical clocks can be caused by their dependence on the stellar birth locations, as the enrichment from AGB stars that have migrated causes the increase in [s/$\alpha$] abundance at a given radius.

From a modelling perspective, reproducing the observed behaviour of chemical clocks remains a significant challenge. Chemical evolution models that successfully match a broad range of abundance ratio trends in the Milky Way (MW) disc often fail to capture the observed rise in [s/$\alpha$] at young stellar ages, with discrepancies that are most pronounced in the inner disc and for second s-process peak elements, such as Ba. In \citetalias{Molero2025}, we explored a number of possible solutions within a multi-zone chemical evolution framework, investigating variations in the disc formation history, the metallicity and mass dependence of AGB nucleosynthetic yields, and the rotational velocity distribution of massive stars. We found that the production of second s-process peak elements in the last 3 Gyr would need to be increased by approximately 50$\%$ to match observations, and while contributions from low-mass ($\sim 1.1\ \rm M_\odot$) AGB stars can partially alleviate this tension, the required enhancement significantly exceeds what current nucleosynthesis models predict (however in agreement with \citealp{Dorazi2009, Maiorca2012}).

In this Letter, we present chemical evolution model predictions similar to those of \citetalias{Molero2025}, comparing them against the high-precision dataset of \citet{casali25}, which consists of 68 \textit{Kepler} red giant stars with asteroseismic ages from individual-mode-frequencies and high-resolution spectroscopy of several neutron-capture elements. In particular, we compare our models with abundance measurements of both first- ([Zr/Ti]) and second-peak ([Ce/Ti]) s-process chemical clocks, to assess whether the nucleosynthetic modifications explored in \citetalias{Molero2025} are able to reproduce the tight observed trends in the disc. The advantage of this dataset with respect to the open clusters in \citetalias{Molero2025} is that it extends to the older regime ($> 7$~Gyr) and provides higher precision measurements than those from \emph{Gaia}-ESO survey. 
The Letter is structured as follows. In Sect~\ref{data} and~\ref{nucleosynthesis}, data sample and Galactic chemical evolution (GCE) models are presented; in Sect.~\ref{results}, the results are discussed. Finally, in Sect.~\ref{conclusions}, we summarise and conclude.

\section{Data sample}
\label{data}
We use the same sample analysed in \citet{casali25}, which is ideally suited for the present work because it combines high-precision asteroseismic ages with detailed optical chemical abundances. The dataset consists of 68 \emph{Kepler} red giant branch stars selected from the RGB catalogue of \citet{miglio21} cross-matched with APOGEE DR17 \citep{majewski17,apogeedr17}. The targets were selected to uniformly sample the low- and high-$\alpha$ sequences over the metallicity interval $-0.8 \leq \mathrm{[Fe/H]} \leq 0.4$ dex. As a result, the sample provides a well-controlled set of disc giants spanning a broad range of age and chemical properties, with the advantage of having precise age constraints from \emph{Kepler} asteroseismology.

As described in \citet{casali25}, complementary high-resolution optical spectra were obtained with HARPS-N \citep[][$R \sim 115,000$]{harps-n} and FIES \citep[][$R \sim 67,000$]{fies} spectrographs, respectively, at the Telescopio Nazionale Galileo (TNG) and Nordic Optical Telescope (NOT), allowing us to extend the chemical characterisation beyond the abundance set available from APOGEE. In particular, the optical wavelength coverage includes the blue spectral region, which is crucial for measuring several neutron-capture species and therefore provides information that is otherwise unavailable, or significantly more limited, in the APOGEE infrared spectra. Atmospheric parameters and chemical abundances were derived through a homogeneous equivalent-width analysis based on DOOp/DAOSPEC \citep{stetson08,cantat14}, MOOG \citep{sneden12}, and MARCS model atmospheres \citep{gustafsson08}, with surface gravities fixed to the seismic values. 

The stellar ages adopted in this work are those used in \citet{casali25}, estimated from Bayesian inference using the AIMS code \citep{aims,rendle19}, APOGEE-DR17 atmospheric parameters, and individual mode frequencies, as in \citet{Montalban21}. 
These ages, which will be published in Montalbán et al (in prep.), have a typical age precision of 9\%. All details on sample selection, observations, abundance analysis, and seismic age determination are present in \citet{casali25}.

\section{Nucleosynthesis}
\label{nucleosynthesis}
Neutron-capture elements are synthesised through both the s- and r-processes, with their relative contributions varying from element to element. The elements analysed in this work, Zr and Ce, are predominantly s-process products, as the s-process accounts for $\sim 82\%$ of Zr and $\sim85\%$ of Ce (at solar metallicity), with the remaining fractions attributed primarily to the r-process and, in the case of Ce, a negligible contribution from the p-process \citep{Prantzos2020}. In this Letter, we focus on the s-process and keep the r-process prescriptions constant among different elements. For a detailed description of those we refer to \citet{Molero2023}.

The s-process operates through three components: the weak, main, and strong s-process. The weak s-process, which is primarily responsible for producing most of the s-process isotopes between Fe and Sr, operates in massive stars (with initial mass $M \gtrsim 8 \rm ~M_{\odot}$) via the neutron source reaction \ce{^{22}Ne(\alpha, n)^{25}Mg}, active mainly during core He- and C-burning \citep{Pignatari2010}. The main and strong s-process components instead occur in low- and intermediate-mass stars {\bf ($1.3 {\rm ~M_{\odot}} \lesssim M \lesssim 8 {\rm ~M_{\odot}}$)} during their AGB phase, where the dominant neutron source is \ce{^{13}C(\alpha, n)^{16}O} \citep{Karakas2010}. In particular, the main s-process is responsible for the production of elements between Sr and Pb excluded (i.e., both the first -- Sr, Y, and Zr -- and second -- Ba, La and Ce -- s-process peaks), while the strong s-process accounts for the third peak, around Pb.

\begin{figure}
    \centering
    \includegraphics[width=0.8\columnwidth]{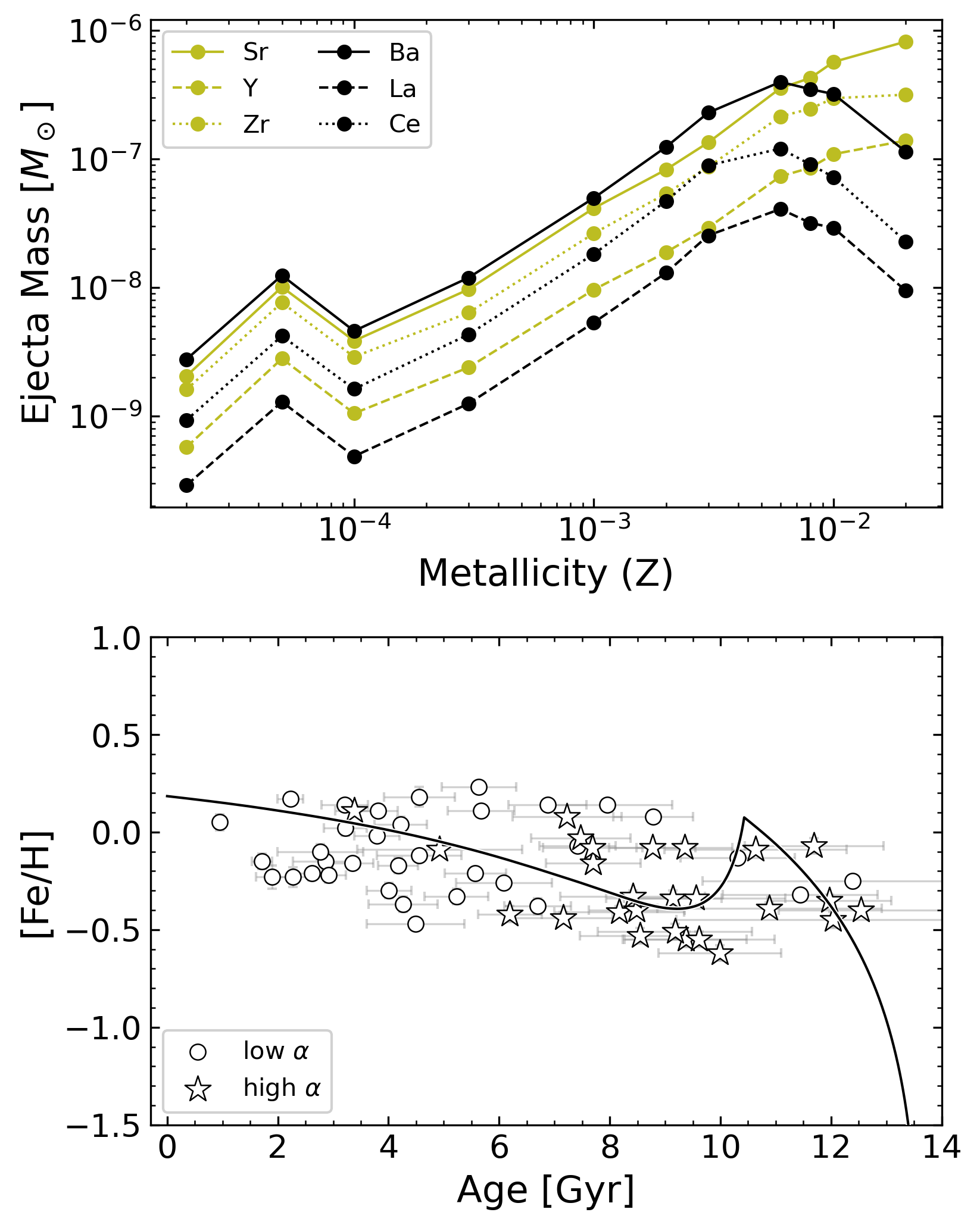}
    \caption{Top panel: FRUITY nucleosynthetic yields for s-process elements of the first peak (Sr, Y, Zr; olive lines) and second peak (Ba, La, Ce; black lines) as a function of metallicity $Z$, for an AGB star of initial mass $\rm 2\ M_\odot$. Bottom panel: predicted [Fe/H] vs. Age relation compared to the \textit{Kepler} red giant stars of \citet{casali25}.}
    \label{fig: AGB yields 2 Msun}
\end{figure}

For massive stars with initial masses $M \rm \geq 13\ M_\odot$  we adopt the rotating stellar yields of set R from \citet{LimongiChieffi2018}, coupled with the distribution of rotational velocities proposed by \citet{pranzos2018}, which favours faster rotation at lower metallicities \citep[see also][]{Romano2019, Rizzuti2021, Molero2024}. 
However, the yields for massive stars are uncertain, as different prescriptions behave differently \citep[e.g.,][]{Frischknecht2016,choplin18,LimongiChieffi2018,Banerjee2019}.
For LIMS with initial masses $1.0 \leq M/\rm M_\odot \leq 8.0$, we adopt yields from the FRUITY database \citep{cristallo09, cristallo11, cristallo15}, which provides an extensive grid spanning 8 progenitor masses (from 1.3 to $6.0 \rm \ M_\odot$) and 12 metallicity values (from $\rm Z=4.8 \times 10^{-5}$ to $\rm Z=2.0 \times 10^{-2}$). We will refer to this model as Model 1. The nucleosynthetic predictions of the FRUITY models for first- and second-peak s-process elements are shown in the top panel of Figure \ref{fig: AGB yields 2 Msun} for an example progenitor of $2\ \rm M_\odot$. While the production of first-peak elements increases with metallicity for $Z \geq 10^{-4}$, the yields of second-peak elements decrease from $Z=7\times10^{-3}$. This behaviour was already noted by, e.g., \citet{gallino98} and \citet{Travaglio1999}. 
Indeed, at higher metallicity, the larger number of iron seed nuclei competing for the available neutron budget lowers the neutron-to-seed ratio, reducing the mean neutron exposure per seed. Since reaching the second peak (A $\sim$ 130) requires more successive neutron captures than the first peak (A $\sim$ 88), this reduced exposure preferentially suppresses second-peak production, while first-peak yields remain comparatively unaffected.
\citet{Ratcliffe2024} tested chemical evolution models with \textit{modified} AGB yields for second-peak s-process elements, replacing the Ba yields at $Z>0.01$ with those at $Z=0.01$. Here, we apply the same modification to the Ce yields, defining this as Model 2. In addition, we test two further models in which the Ce yields at the highest metallicity steps are increased by factors of 1.5 and 2.0, referred to as Model 3 and Model 4, respectively.

Finally, the Age-metallicity relation predicted by the models is shown in the bottom panel of Figure \ref{fig: AGB yields 2 Msun}.

\section{Results and Discussion}
\label{results}
We compare our GCE model predictions with the high-precision chemical clock data of \citet{casali25}, focusing on [Zr/Ti], representative of the first s-process peak, and [Ce/Ti], representative of the second peak. The choice of Ti as the reference $\alpha$-element is driven by the observational dataset adopted, in which the [Zr/Ti] and [Ce/Ti] vs. Age relations are the tightest among the available chemical clocks, with a scatter of $s = 0.01\ \rm dex$ and of $s = 0.08\ \rm dex$ for Zr and for Ce, respectively (see Figure 6 of \citealp{casali25}). We do, however, acknowledge that Ti is problematic from the theoretical point of view, as it is systematically underproduced by all CC-SN yield sets and as a consequence GCE models (e.g., \citealp{pranzos2018, Kobayashi2020, Jost2025, Pepe2026}). Because of that, we apply a metallicity-independent scaling factor of 4 to the CC-SN Ti yields adopted. Such an adjustment is clearly a phenomenological rescaling adopted so that Ti can be used as the reference $\alpha$-element. To verify that our conclusions are not modified by this choice, we extended the analysis also to the $\alpha$-elements Si and Ca, which do not require corrections, but that in the adopted dataset show less tight chemical-clocks relations. Results for those elements are reported in the Appendix \ref{s/ca s/Si}.

\begin{figure}
    \centering
    \includegraphics[width=0.8\columnwidth]{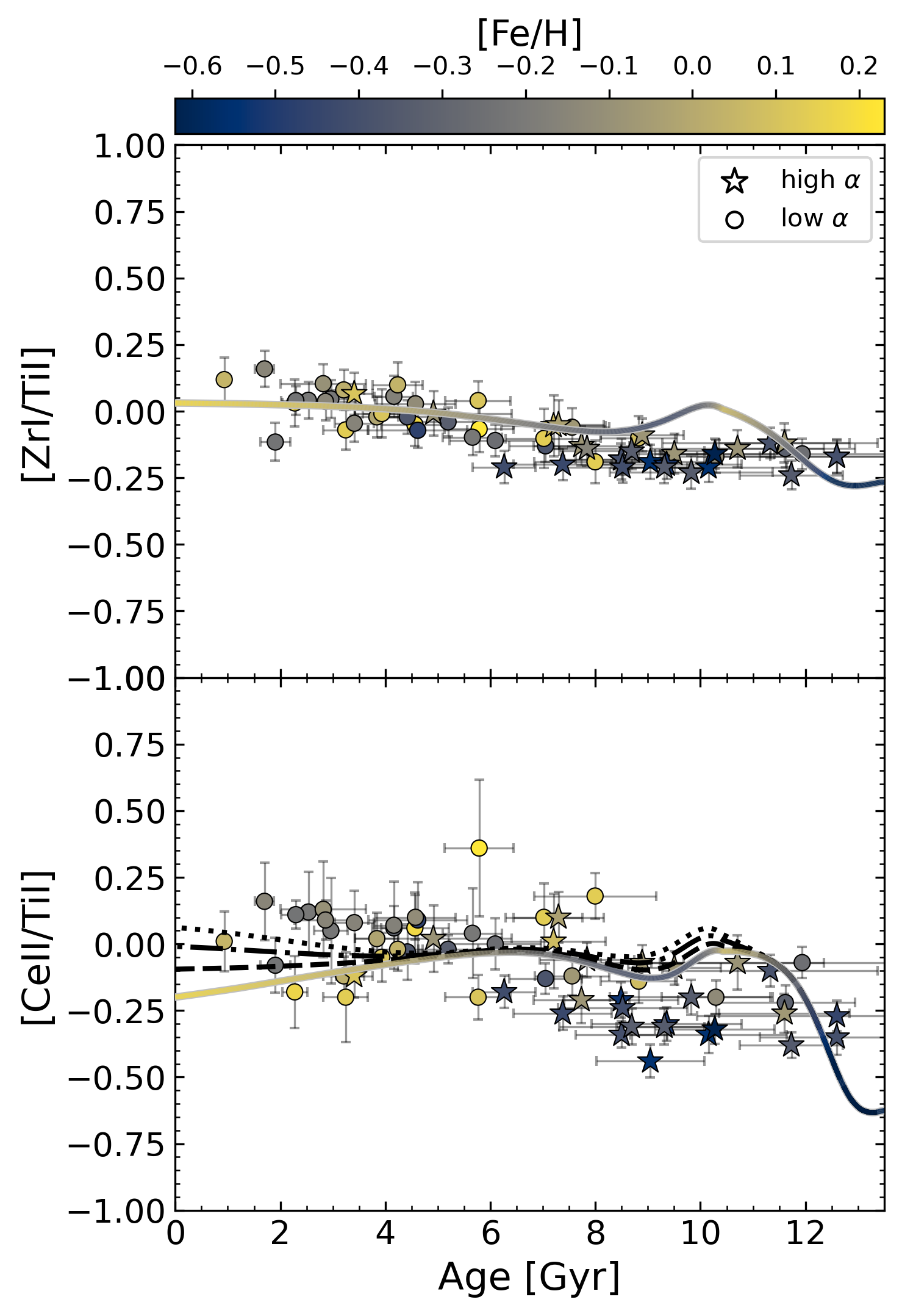}
    \caption{[Zr/Ti] and [Ce/Ti] as a function of stellar age for the 68 \emph{Kepler} red giant stars of \citet{casali25}, colour-coded by [Fe/H]. Stars belonging to the high- and low-$\alpha$ sequences are indicated by stars and circles, respectively. The two-infall model (Model 1) predictions are shown as solid coloured lines, colour-coded in the same way as the data. The other lines in the [Ce/Ti] panel show the effect of different nucleosynthetic prescriptions for AGB stars: dashed line is Model 2, dash-dotted line is Model 3, dotted line is Model 4 (see text for details).}
    \label{fig: chemical clocks vs. Age}
\end{figure}

The observed trends of the [Zr/Ti] and [Ce/Ti] chemical clocks span a wide range of stellar ages, from $\sim$1 Gyr to $\sim$13 Gyr (see Fig. \ref{fig: chemical clocks vs. Age}), extending to ages older than those probed by the open cluster sample of the \emph{Gaia}-ESO survey previously adopted in \citetalias{Molero2025} and being also larger in number, allowing us to have a more stringent test of the model across the entire chemical evolution history. Both chemical clocks show the expected increasing trend with decreasing stellar age, reflecting the delayed enrichment of s-process elements relative to $\alpha$-elements. The high-$\alpha$ sequence is predominantly composed of old, metal-poor stars, while the low-$\alpha$ sequence spans intermediate to young ages at higher metallicities, consistent with the picture of an early, rapid enrichment phase followed by a more quiescent evolution of the thin disc. There are also a few old ($\sim 10-11$ Gyr) metal-poor low-$\alpha$ stars. These stars suggest an earlier formation scenario for the MW thin disc compared to current estimates of $\sim 8-9 $ Gyr for the onset of thin-disc formation \citep[see also][for further detail]{nepal24}.

The two chemical clocks are compared with the results of a two-infall chemical evolution model (see \citealp{spitoni2021} for a description), with the different nucleosynthetic prescriptions described in the previous section. For [Zr/Ti], Model 1 (solid coloured line in Fig.~\ref{fig: chemical clocks vs. Age}) predicts an increase towards younger ages, though with a slight shallower slope than observed. For [Ce/Ti], on the other hand, the model tends towards a decreasing trend at young ages, in particular from $\rm Age \lesssim 6\ \rm Gyr$, in disagreement with the observations. Furthermore, for both clocks the model shows a local peak at $\sim 10-11\ \rm Gyr$, that is not observed in the data, in particular in the tight age sequence traced by [Zr/Ti]. The presence of this feature in the models may indicate that the relative nucleosynthetic timescales of the elements considered differ, at early times, from those implied by our adopted yields, and/or that the assumed history of star formation requires modifications. We explored the first possibility, and the corresponding results are reported and discussed in the Appendix \ref{rotational distributions}.

In Fig. \ref{fig: chemical clocks vs. FeH}, the same model is shown as a function of [Fe/H]. The agreement with the high-$\alpha$ sequence is generally satisfactory, whereas the low-$\alpha$ sequence is less well reproduced. In particular, following the dilution of the interstellar medium (ISM) at the onset of the second infall episode (which drives the model track towards lower [Fe/H]) the model re-enriches in metallicity but fails to reach the [Zr/Ti] and [Ce/Ti] values observed at intermediate [Fe/H] values, underestimating the abundance ratios of the low-$\alpha$ sequence across the range $-0.3 \lesssim \rm [Fe/H] \lesssim 0.0$. This suggests that, if the two-infall scenario is correct, the re-enrichment phase following the onset of the second infall should be accompanied by a stronger production of s-process elements than currently predicted, pointing once more to a deficit in the nucleosynthetic yields at the metallicities characteristic of the young thin disc.

The deficit in the production of second-peak s-process material may originate from the lower yields predicted for AGB stars at high metallicity. When this deficit is modified following the prescription of Model 2 (the dashed line), the decrease in [Ce/Ti] vs. Age predicted by the chemical evolution model is indeed reduced for Age $\lesssim\ 4\ \rm Gyr$. Some discrepancies, however, still remain. In particular, because the modification to the s-process yields is introduced above a given metallicity threshold \textbf{($Z > 0.01$)}, it affects the shape of the curve not only at late times, but also at earlier epochs. Specifically, the peak at Age $\simeq 10\ \rm Gyr$ becomes more pronounced, as it corresponds to the end of the first infall phase, when the ISM has already been enriched to relatively high metallicities (see the [Fe/H] colour bar in Fig. \ref{fig: chemical clocks vs. Age}). Moreover, although the decrease predicted by the model at young ages is partially corrected, Model 2 still fails to reproduce the observed increase of the [Ce/Ti] chemical clock, as the predicted decrease remains too pronounced. This behaviour is similarly evident in the [Ce/Ti] vs. [Fe/H] relation: an increase in the Ce yields towards higher metallicities improves the agreement with the observations at intermediate [Fe/H] values, but the model still underestimates the observed trend. 

In agreement with the estimated deficit in second-peak s-process material reported in \citetalias{Molero2025}, we test two additional models in which the Ce yields from AGB stars are boosted at the highest metallicity steps by factors of 1.5 and 2.0, referred to as Models 3 and 4 (the dash-dotted line and the dotted line, respectively). The results are shown in Figs. \ref{fig: chemical clocks vs. Age} and \ref{fig: chemical clocks vs. FeH}. The modified Ce yields of those models have a modest effect on the [Ce/Ti] vs. Age, slightly increasing the predicted abundances at young ages and bringing the models into marginally better agreement with the data at Age $\lesssim\ 4\ \rm Gyr$. However, none of the models is able to fully reproduce the most Ce-enhanced young stars, suggesting that a simple rescaling of the high-metallicity AGB yields is insufficient to account for the observed Ce enrichment at recent times. In the [Ce/Ti] vs. [Fe/H] plane, the boosted Ce yields of Models 3 and 4 shift the model predictions towards higher [Ce/Ti] at intermediate to high [Fe/H], improving the agreement with the data, but without reproducing the most enhanced [Ce/Ti] stars at [Fe/H] $\simeq-0.2\ \rm dex$.


\begin{figure}
    \centering
    \includegraphics[width=0.8\columnwidth]{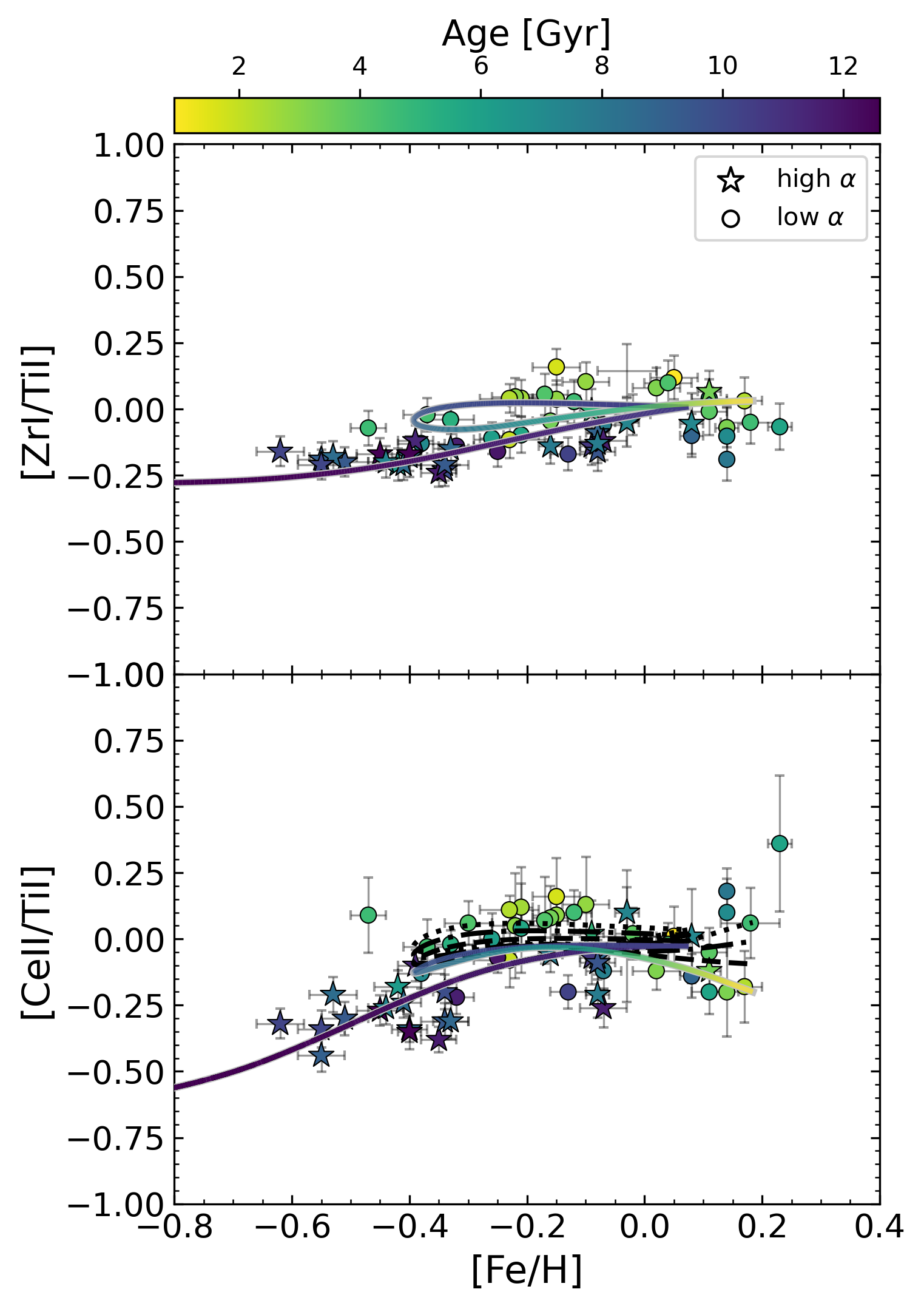}
    \caption{Same as Fig. \ref{fig: chemical clocks vs. Age} but for [Zr/Ti] and [Ce/Ti] as a function of [Fe/H], colour-coded by the age of the stellar population.}
    \label{fig: chemical clocks vs. FeH}
\end{figure}

\begin{figure}
    \centering
    \includegraphics[width=0.80\columnwidth]{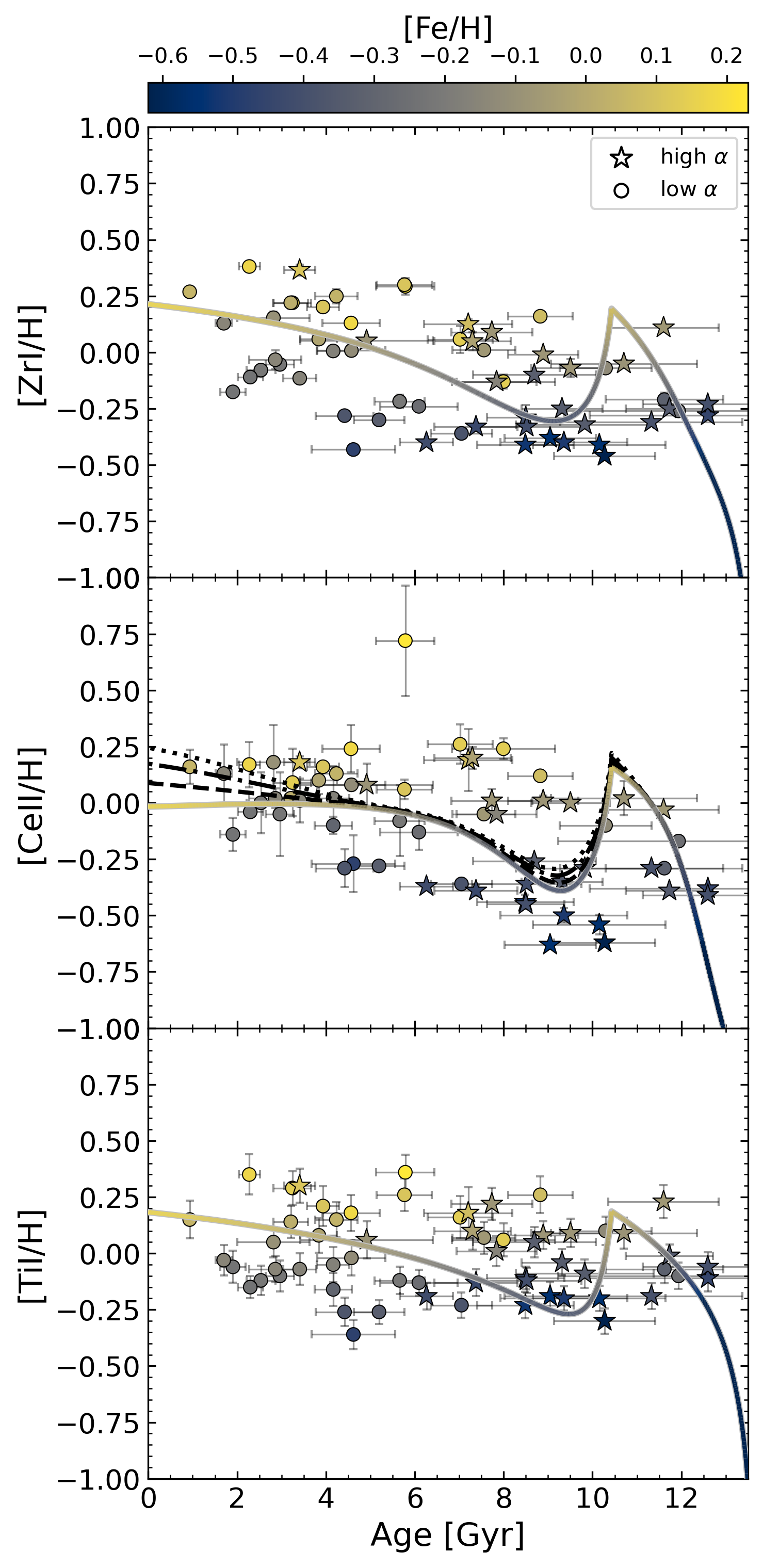}
    \caption{Same as Fig. \ref{fig: chemical clocks vs. Age} but for [Zr/H], [Ce/H], and [Ti/H] as a function of stellar age, colour-coded by [Fe/H].}
    \label{fig: [X/H] vs. Age}
\end{figure}

\begin{figure}
    \includegraphics[width=0.80\columnwidth]{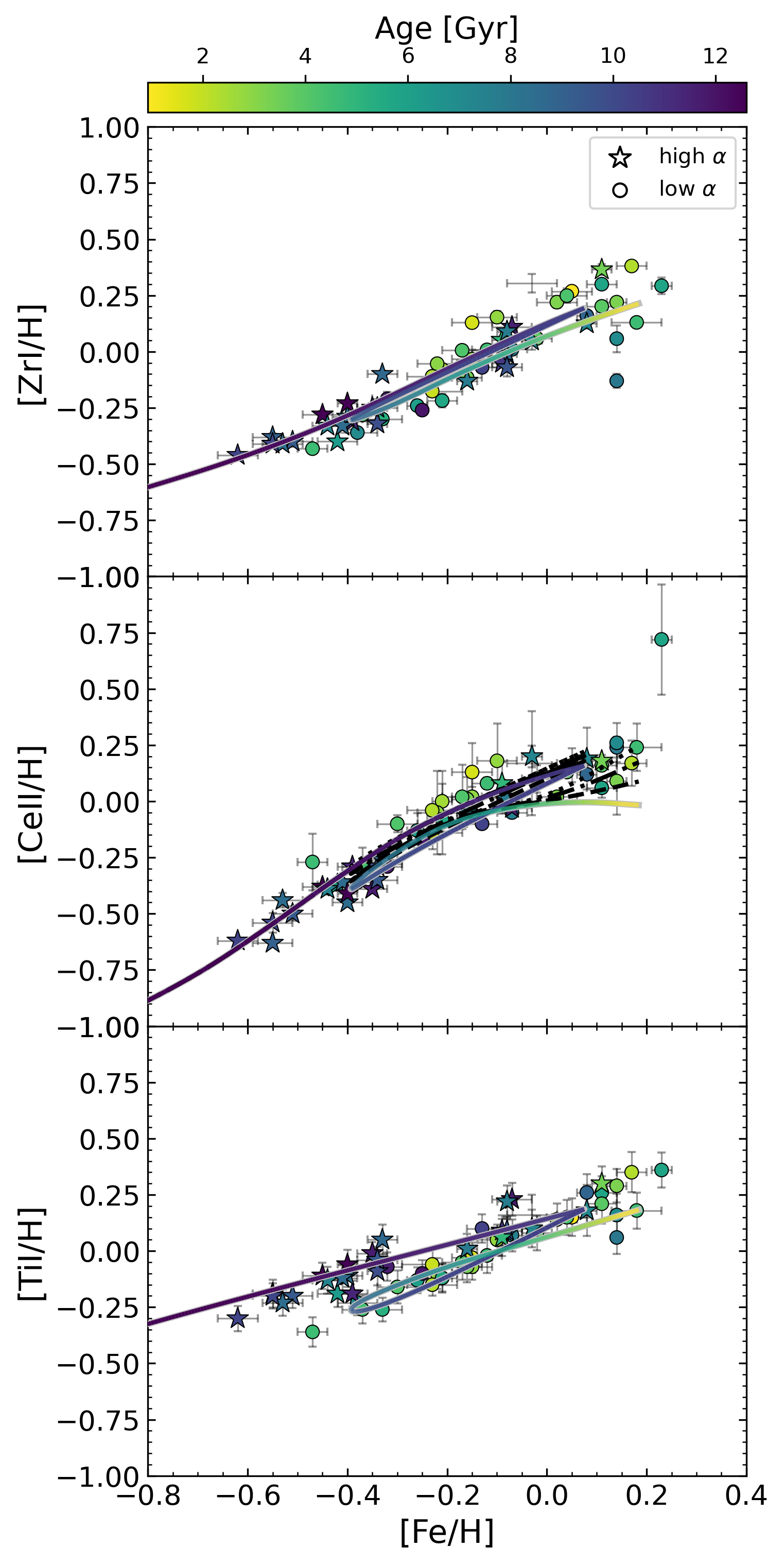}
    \caption{Same as Fig. \ref{fig: chemical clocks vs. Age} but for [Zr/H], [Ce/H], and [Ti/H] as a function of [Fe/H], colour-coded by the age of the stellar population.}
    \label{fig: [X/H] vs. [Fe/H]}
\end{figure}

In Figs. \ref{fig: [X/H] vs. Age} and \ref{fig: [X/H] vs. [Fe/H]}, we show the individual abundances [Zr/H], [Ce/H], and [Ti/H] as a function of stellar age and [Fe/H]. All three elements display the expected increasing trend with decreasing age, reflecting the progressive enrichment of the ISM over time. The relation is a bit tighter for Ti than for the neutron capture elements and characterised by a less steep increase. The model reproduces the overall shape of the [Ti/H] and of the [Zr/H] vs. Age relations reasonably well across the full age range, but for [Ce/H] the model gets flatter towards younger ages, as a consequence of the deficit in second-peak s-process material already noted in the chemical clock ratios. The characteristic local peak at $\mathrm{\sim 10-11\ \rm Gyr}$, driven by the high-metallicity enrichment at the end of the first infall phase, is visible in the model predictions for all the elements, but is not clearly seen in the data. The boosted Ce yield models (Models 2, 3 and 4) partially compensate the deficit at young ages, but they still fail to reproduce the observed increase of [Ce/Ti] at recent times, as discussed above. It is interesting to note that, from a comparison of the two panels in the figure, the model agreement with observations is better in metallicity space than in age space. This is apparent both from the [X/H] vs. [Fe/H] panels, where the model tracks closely follow the observed sequences, and from the colour-coded model lines in the [X/H] vs. Age panels. The overall age evolution is in fact well reproduced by the models, but compared to the metallicity space, the age space is characterised by a larger observed scatter, with stars of similar age spanning a wide range of [X/H] values and metallicities. 
Indeed, a large scatter of [X/H] vs. Age is also naturally produced from the inhomogeneity of the chemical enrichment in chemodynamical simulations \citep{kobayashi2011}. Moreover, it might be due to the intrinsic spread in birth radii of stars currently observed in the solar neighbourhood. The effect of stellar migration, currently not accounted for in our model, is to broaden the observed [X/H] vs. Age relations at a fixed Galactocentric distance more than what is seen in the corresponding [X/H] vs. [Fe/H] relations. In fact, stars observed today in the solar neighbourhood were born at a range of Galactocentric radii, each with its own chemical enrichment history. 
Migrated stars tend to scatter predominantly along the age axis rather than the metallicity axis. Stars with similar [Fe/H] may show comparable [X/H] patterns regardless of their formation site. In contrast, stars of similar age can span a wide range of [X/H] depending on their birth radius, which increases the apparent scatter in age space (\citealp{Wang2013, Minchev2013, Spitoni2015}).
A comprehensive treatment of radial migration and its impact on the chemical clocks relations and on the evolution of s-process elements as a function of stellar age is the subject of ongoing work (Molero et al., in preparation).

As discussed in \citetalias{Molero2025}, adjustments to AGB yields at high metallicities must be approached with caution, for several reasons. First, the high metallicities reached by the chemical evolution model at young ages in the solar neighbourhood are also reached in the inner disc at older ages, as a consequence of the inside-out formation scenario, which predicts stronger and more efficient star formation in the inner regions of the MW disc. Therefore, increasing AGB yields at high metallicity would not only steepen the chemical clock trends towards younger ages in the solar neighbourhood, but would also shift the entire chemical evolution track of the inner disc to higher [s/$\alpha$] values over a wide range of ages, producing strong disagreement with open cluster observations at $R_{\rm GC} \lesssim 7\ \rm kpc$. Second, within the two-infall model of disc formation, the high metallicities we are attempting to correct are also reached at very early times ($\sim 10-11$ Gyr ago), driven by the first strong infall event. An increase of the yields at high metallicities would therefore produce an excess peak in the [s/$\alpha$] and [s/H] vs. Age trends at the end of the first infall phase, which is not observed in the data. Third, if high-metallicity AGB progenitors were to enhance their s-process production, it is unclear why this effect should selectively boost second-peak elements while leaving first-peak elements unaffected. Any enhancement that affects the second peak would simultaneously alter the first-peak chemical clocks, worsening their agreement with observations. This tension is further compounded by the fact that the production of second-peak s-process elements from AGB stars is theoretically expected to decrease with increasing metallicity (see discussion in Sec.~\ref{nucleosynthesis}), as also supported by observations of decreasing [heavy-s/light-s] ratios with increasing [Fe/H] in Ba stars, in agreement with current nucleosynthesis models of low-mass AGB stars (\citealp{Cseh2018}). 


\section{Summary and conclusions}
\label{conclusions}
In this Letter, we have presented Galactic chemical evolution (GCE) model predictions for the [Zr/Ti] and [Ce/Ti] chemical clocks, comparing them against the high-precision sample of 68 \textit{Kepler} red giant branch stars from \citet{casali25}, which combines asteroseismic ages from individual-mode-frequencies with high-resolution optical spectroscopy. This dataset extends the age baseline beyond that accessible with the open clusters dataset of the \emph{Gaia}-ESO survey previously adopted in \citetalias{Molero2025} (see also \citealp{Magrini2023}), providing a more stringent test for the models across the entire disc evolution. Our main results can be summarised as follows:

\begin{itemize}
    \item The baseline GCE model (Model 1) satisfactorily reproduces the [Zr/Ti] chemical clock and the high-$\alpha$ sequence in both age and metallicity space, reflecting a well-constrained enrichment history for first s-process peak elements. For [Ce/Ti], however, the model predicts a decreasing trend at young ages (Age $\lesssim$ 6 Gyr) in disagreement with the observed rise, pointing to a persistent deficit in second s-process peak production during the last Gyrs of thin disc evolution.
    \item Modifying the Ce yields of high-metallicity AGB stars following the prescription of Model 2, or increasing them by factors 1.5 and 2.0 (Models 3 and 4), only partially alleviates this tension. While these modifications improve the agreement at intermediate [Fe/H] and young ages, none of the models is able to fully reproduce the most Ce-enhanced young stars, indicating that a simple rescaling of the high-metallicity AGB yields is insufficient.
    \item Both chemical clocks show a characteristic local peak in the model predictions at $\sim$10-11 Gyr, associated with the high-metallicity enrichment at the end of the first infall phase, which is not clearly seen in the data. This discrepancy persists across all nucleosynthesis prescriptions tested and may point to limitations in the adopted disc formation scenario.
    \item The models reproduce the observed abundance trends more tightly in metallicity space. The additional scatter observed in the [X/H] vs. Age relations can be attributed to the inhomogeneity of the chemical enrichment and the spread in birth radii of stars currently observed in the solar neighbourhood, a consequence of radial migration that is not accounted in the adopted framework.
\end{itemize}

Our analysis shows that modifications of AGB yields at high metallicity must be treated with caution. In particular, the high-metallicity regime relevant to the solar neighbourhood is also probed at earlier times in the inner disc due to inside-out Galaxy formation, meaning that any increase in AGB yields would simultaneously change chemical clock trends locally and degrade the agreement with open cluster abundances at small Galactocentric radii. Moreover, in two-infall models, high metallicities are reached during the early rapid infall phase, so such modifications of AGB yields would produce unobserved features in the age-abundance relations. Finally, there is no clear nucleosynthetic justification for selectively enhancing second-peak s-process elements without affecting first-peak species, especially given that higher metallicity is expected to reduce neutron-to-seed ratios and therefore suppress second-peak production. Overall, these considerations argue against ad hoc high-metallicity yield enhancements and highlight the need for physically self-consistent chemical evolution models.

In conclusion, reproducing the behaviour of second s-process peak chemical clocks in the young thin disc remains an open challenge. The origin of the missing s-process enrichment at high metallicity and young ages might be connected to uncertainties in AGB nucleosynthesis that are beyond a simple rescaling, and a self-consistent treatment of radial migration within the GCE framework may further contribute to reconciling models with observations. With this Letter we aim to stimulate further work in stellar yields and Galactic chemical evolution, providing possible avenues for future investigation and highlighting the importance of high-precision data sets for constraining and refining Galactic chemical evolution models.

\section*{Acknowledgements}

GC and MM thank the anonymous referee for the useful comments that improved the quality of this Letter. GC and MM thank Dr. Josefina Montalbán for providing access to data that are not yet published and that will be presented in Montalbán et al. (in preparation).
MM acknowledges support by the Deutsche Forschungsgemeinschaft
(DFG, German Research Foundation) – Project-ID 279384907 – SFB 1245. GC acknowledges
support from the European Research Council Consolidator Grant funding scheme (project ASTEROCHRONOMETRY, G.A. n. 772293, \url{http://www.asterochronometry.eu}. 
This article is based on observations made with the Italian Telescopio Nazionale Galileo (TNG) operated on the island of La Palma by the Fundación Galileo Galilei of the INAF (Istituto Nazionale di Astrofisica) at the Spanish Observatorio del Roque de los Muchachos of the Instituto de Astrofisica de Canarias. 
This article is also based on observations made with the Nordic Optical Telescope (NOT), owned in collaboration by the University of Turku and Aarhus University, and operated jointly by Aarhus University, the University of Turku and the University of Oslo, representing Denmark, Finland and Norway, the University of Iceland and Stockholm University at the Observatorio del Roque de los Muchachos, La Palma, Spain, of the Instituto de Astrofisica de Canarias.
GC and MM acknowledge A. Miglio, A. Bragaglia, M. Matteuzzi, K. Brogaard, A. Stokholm, V. Grisoni, M. Tailo, and E. Willett for their contributions to the observing proposals (A45TAC\_22, A47TAC\_9, P65-006) at the TNG and NOT telescopes.
Computations, provided by Dr. J. Montalbán and described in this paper, were performed using the University of Birmingham's BlueBEAR HPC service, which provides a high-performance computing service to the university’s research community. See \url{http://www.birmingham.ac.uk/bear} for more details.

\section*{Data availability}
The data sample used in this work is available in \citet{casali25}. All other data will be made available upon reasonable request to the authors.



\bibliographystyle{mnras}
\bibliography{Bibliography} 




\appendix

\begin{appendix}

\section{The chemical-clocks [Zr/Ca], [Ce/Ca] and [Zr/Si], [Ce/Si].}
\label{s/ca s/Si}

To verify that our conclusions are not, at least partially, modified by the choice of the adopted reference $\alpha$-element, we repeated the analysis with other two $\alpha$-elements available in \citet{casali25}, Ca and Si, which do not suffer the same issues previously discussed for Ti and therefore do not need any \textit{ad hoc} correction. However, the observed chemical clocks [Zr/Ca] and [Zr/Si] vs. Age relations are not as tight as those based on Ti (with scatter $s = 0.03\ \rm dex$ and $s = 0.12\ \rm dex$, respectively), while the clocks [Ce/Ca] and [Ce/Si] are as tight as those based on Ti ($s = 0.08\ \rm dex$) and thus provide an optimal independent test. Moreover, Si as the reference $\alpha$ element adopted in our previous work, \citetalias{Molero2025}, so its comparison against the extended dataset used here represents a natural continuation of our analysis.

\begin{figure}
    \centering
    \includegraphics[width=1\columnwidth]{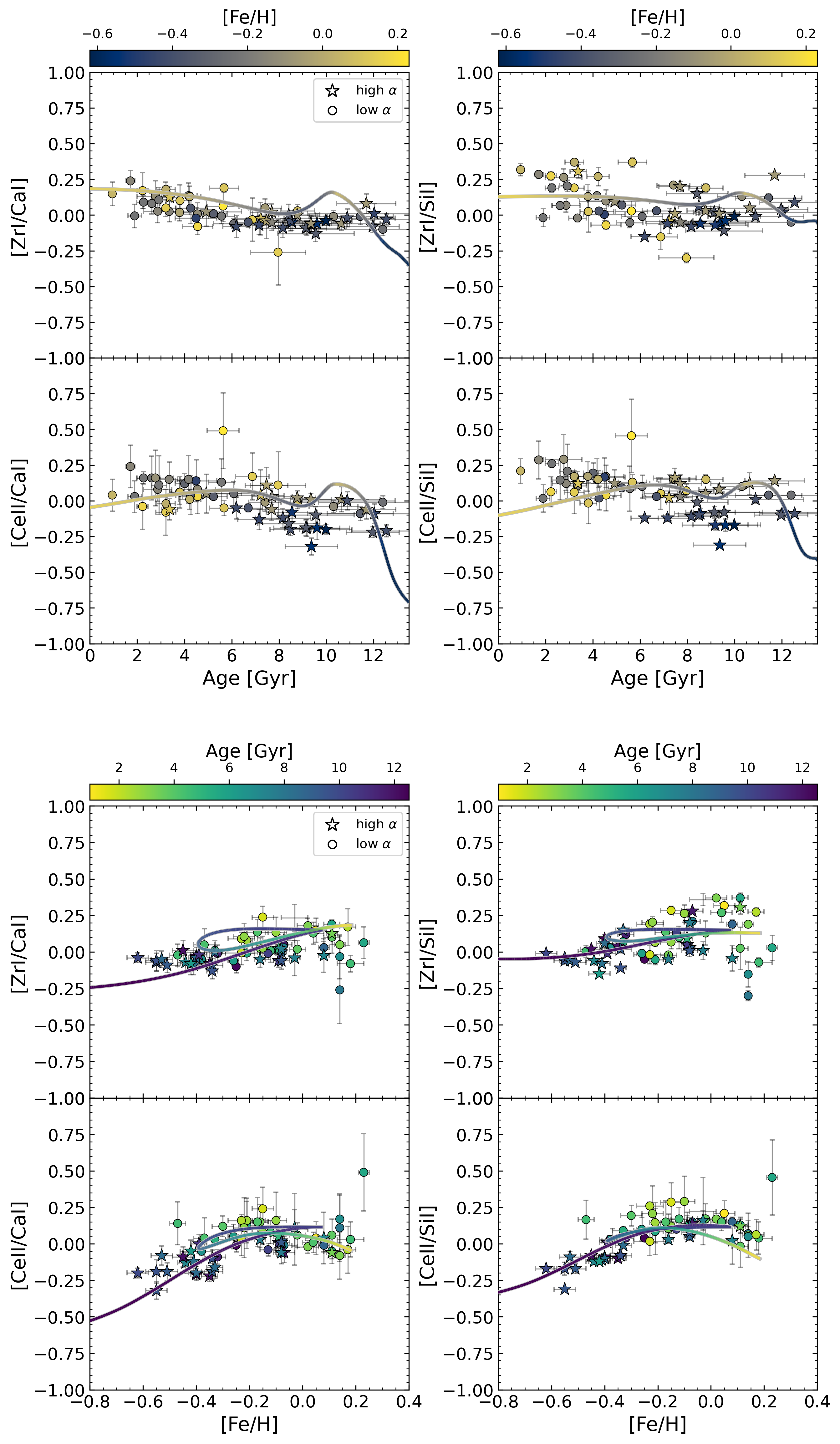}
    \caption{Top panels: as Figure \ref{fig: chemical clocks vs. Age}, and bottom panels: as Figure \ref{fig: chemical clocks vs. FeH}, but for the [Zr/Ca], [Ce/Ca] (left) and [Zr/Si], [Ce/Si] (right), for Model 1.}
    \label{fig:Si and Ca chemical clocks}
\end{figure}

In Figure \ref{fig:Si and Ca chemical clocks} we show the [s/$\alpha$] relations vs. Age and vs. [Fe/H], for both the observational data and our Model 1 results. Compared with the corresponding relations obtained using Ti, we find no major differences that would affect our conclusions. Observationally, both chemical clocks rise towards younger ages, and this rise is better reproduced for Zr than for Ce. The model also produces a peak at young ages, already present in the Ti case and likewise seen here for both Ca and Si, which is not evident in the data (see next section for further discussion). In the metallicity plane, the [Ce/Ca] and [Ce/Si] values of the intermediate [Fe/H] stars are underproduced by the model, exactly as found for Ti, whereas the most metal-rich stars are well reproduced. In the case of Ca, the only difference with respect to the [s/Ti] vs. [Fe/H] relations concerns the most metal-poor stars of the high-$\alpha$ sequence, which appear slightly underproduced when Ca is used as the reference elements. In the case of Si, those stars are well reproduced by the model, but the high-metallicity ones are not. In particular in the case of the [Zr/Si], for which the scatter is however observed to be large even at such high metallicity.



In all cases, however, the conclusions drawn from the Ti-based relations concerning the underestimation of the [Ce/$\alpha$] chemical clocks remain valid and independent of the adopted $\alpha$-element.

\section{The effect of massive star rotational distributions.}
\label{rotational distributions}

All the chemical clock relations predicted by the GCE models show the presence of a peak at $\sim10-11\ \rm Gyr$ not present in the data, in particular not in the very tight [Zr/Ti] vs. Age relation. The relative timing of production of the two different elements can be partially responsible for the presence of such a peak and to its shape. In particular, rotation of massive stars does affect the shape of the peak, but a higher rotational velocity does not necessarily accentuate it.

In general, faster rotation of massive stars enhances the production of all the elements considered here. Moving from 0 to 150 km/s, the enhancement is larger for first-peak than for second-peak s-process elements (see e.g., Figure 2 and 3 of \citealp{Rizzuti2019}; Figure 4 of \citealp{Molero2023}), but at the highest velocity of 300 km/s it becomes comparably strong for both.

\begin{figure}
    \centering
    \includegraphics[width=1\columnwidth]{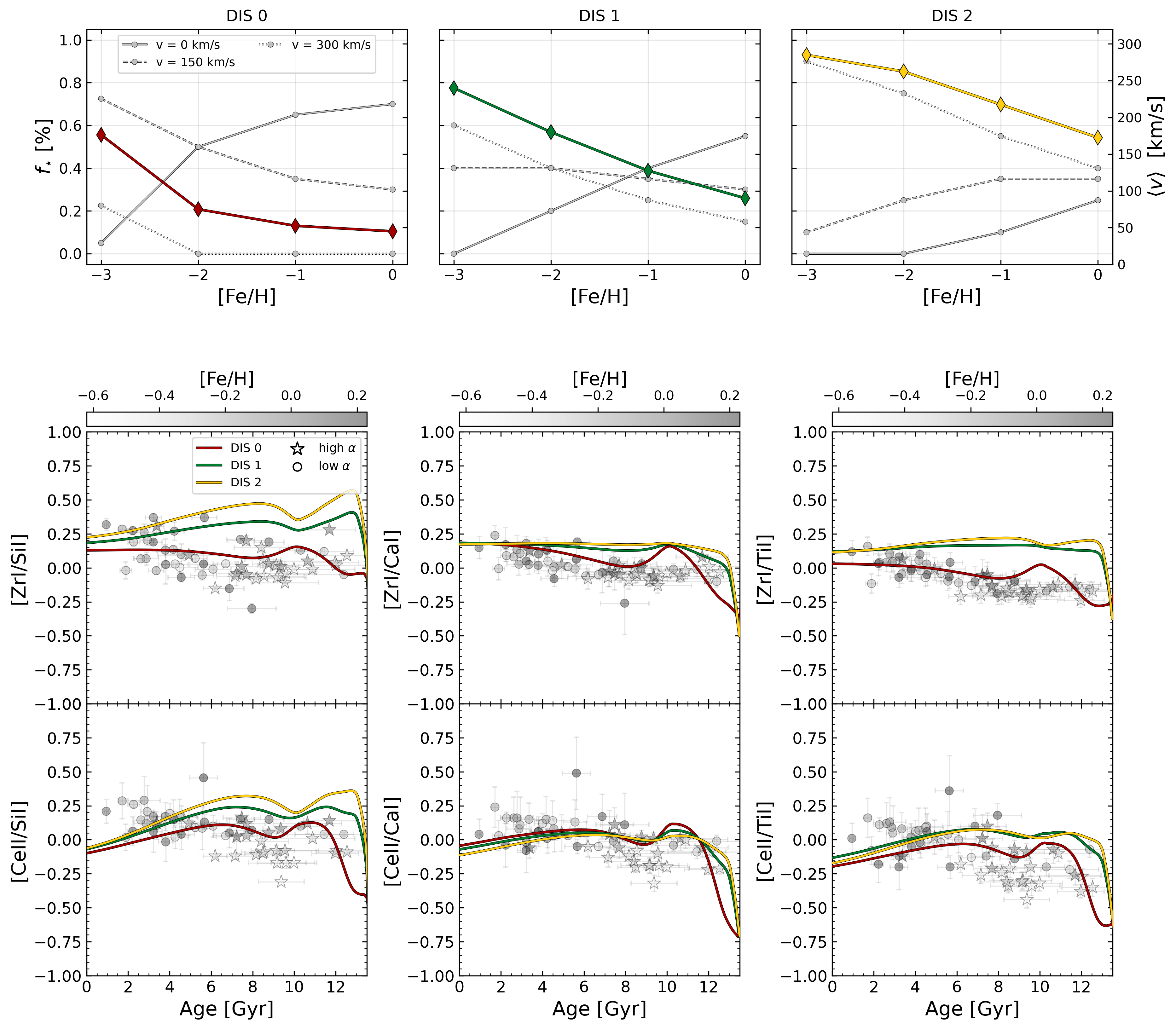}
    \caption{Top panel: Distribution of rotational velocities adopted for the massive star yields. The grey curves show the fraction of massive stars assigned to each rotational velocity (0 km/s solid, 150 km/s dashed, and 300 km/s dotted) as a function of metallicity. Coloured lines/diamonds (right-hand axis) give the corresponding mean rotational velocity. From left to right the distributions become progressively dominated by fast rotators, and in all cases the fraction of fast rotators decreases with increasing metallicity. Bottom panel: as Figure \ref{fig: chemical clocks vs. Age}, but for [Zr/Si], [Ce/Si] (left), [Zr/Ca], [Ce/Ca] (centre), and [Zr/Ti], [Ce/Ti] (right), for predictions from the models correspondent to the three distribution of rotational velocities of the top panels.}
    \label{fig: DIS2}
\end{figure}

It is worth investigating to what extent higher rotational velocities could smooth out the predicted peak. To this end, we ran two additional models adopting alternative distributions of rotational velocities for massive stars which, relative to the baseline distribution adopted in Models 1, 2, 3, and 4, assume a larger fraction of fast rotators at all metallicities. These distributions are shown in Figure \ref{fig: DIS2} (top panel), together with the corresponding [s/$\alpha$] vs. Age relations (bottom panel). As we see, increasing the rotation progressively reduces the peaks in the chemical clock relations for [s/Ca] and [s/Ti], with the fastest distribution yielding to the flattest trends. It is important to note that this smoothing is not driven by a suppression of the peak in the individual [X/H] relations, as the peak and dip features due to the two-infall scenario is retained by all the elements. What changes is that, with increasing rotational velocity, we not only produce more s-process and $\alpha$-elements, but the s-process material is increasingly released on prompt timescales, similar to those of Ca and Ti (see the green and yellow curves at 11-13 Gyr). As a result, the [X/H] vs. Age relations of the four elements become nearly parallel (especially at older ages and for Zr and Ca), so their common shape cancels in the [s/$\alpha$] ratio and the peak is smoothed out. It is important however to note that, even where the peak is smoothed out the agreement with the data is lost in almost all panels.

We therefore conclude that the rotational velocities of massive stars do have an effect on the peak observed at older ages. Whether this smoothing is achieved, however, also depends on the reference $\alpha$-elements: while the [s/Ca] and [s/Ti] peaks are progressively flattened by faster rotation, the peak in [s/Si] persists even for the most extreme distributions. We note that, even if the observed [s/Si] relations are more scattered, they do not show a clear peak. In any case, the possibility that the smoothing of this feature is entirely due to the massive star yields can probably be ruled out: a full cancellation of the peak is obtained only with extreme models, which are disfavoured both on independent grounds and by the chemical clocks relations themselves.

\end{appendix}


\label{lastpage}
\end{document}